\documentstyle[12pt]{article}
\begin{document}

\title{{\small \noindent hep-th/9604139\hfill USC-96/HEP-B3}\bigskip\\
Supersymmetry, p-brane duality and\\
hidden spacetime dimensions}
\date{}
\author{Itzhak Bars\bigskip\thanks{%
Research supported by DOE grant No. DE-FG03-44ER-40168} \\
%EndAName
Department of Physics and Astronomy\\
University of Southern California\\
Los Angeles, CA 90089-0484, USA}

\maketitle

\begin{abstract}
A global superalgebra with 32 supercharges and all possible central
extensions is studied in order to extract some general properties of duality
and hidden dimensions in a theory that treats $p$-branes democratically. The
maximal number of dimensions is 12, with signature (10,2), containing one
space and one time dimensions that are hidden from the point of view of
perturbative 10-dimensional string theory or its compactifications. When the
theory is compactified on $R^{d-1,1}\otimes T^{c+1,1}$ with $d+c+2=12,$
there are isometry groups that relate to the hidden dimensions as well as to
duality. Their combined intersecting classification schemes provide some
properties of non-perturbative states and their couplings.
\end{abstract}
\bigskip
%%%%%%%%%%%%%%%%%%%
% The following constructs ``\Sp, \Sb, \stackunder " 
% are used in Table I and eq.(1).
% they have been modified from tcilatex.tex
%
\def\Let@{\relax\iffalse{\fi\let\\=\cr\iffalse}\fi}
\def\vspace@{\def\vspace##1{\crcr\noalign{\vskip##1\relax}}}
\def\multilimits@{\bgroup \vspace@\Let@
 \lineskip 2pt
 \lineskiplimit\lineskip
 \vbox\bgroup\ialign\bgroup\hfil$\scriptstyle{##}$\hfil\crcr}
\def\Sb{_\multilimits@}
\def\endSb{\crcr\egroup\egroup\egroup}
\def\Sp{^\multilimits@}
\let\endSp\endSb
\def\stackunder#1#2{\mathrel{\mathop{#2}\limits_{#1}}}
%%%%%%%%%%%%%%

\section{Introduction}

The discovery of string dualities have led to the idea that there is a more
fundamental theory than string theory that manifests itself in different
forms in certain regimes of its moduli space. The several familiar string
theories (type-I, type-II, heterotic) may be regarded as different starting
points for perturbative expansions around some vacuua of the fundamental
theory, in analogy to perturbative expansions around the different vacuua of
spontaneously broken gauge theories. A lot of evidence has accumulated by
now to convince oneself that the different versions of $D=10$ superstrings
and their compactifications are related to each other non-perturbatively by
duality transformations \cite{dualities}. Furthermore, there is evidence
that the non-perturbative theory is hiding higher dimensions \cite{witten}-%
\cite{townsend11} (in the form of ``M-theory'' \cite{townsend11}-\cite
{horavawitten}) and that it is related to various $p$-branes \cite{townsend}
and $D$-branes \cite{pol}.

In order to explore the fundamental theory it is desirable to find a common
ground which is valid for the many versions of the theory and that is
independent of the details of the language used to describe the theory. In
search of such a common ground in various dimensions I will explore a
supersymmetry algebra that has 32 fermionic generators and all possible
central extensions \cite{townsend}. The supersymmetry is not necessarily
exact; it may be broken by central extensions that are included in the
algebra. The spirit is similar to charge or current algebras used in the
1960's for weak and strong interactions, avoiding complexities due to the
details of the theory. The basic assumption that we make is that the
superalgebra is valid in the sense of a (broken) dynamical symmetry that
applies to the full theory at the level of matrix elements for (broken)
supermultiplets. By studying the isometries of the superalgebra, including
the central extensions, many of the features of duality may be displayed
while some new features become apparent, including the following:

\begin{enumerate}
\item  The central extensions (as well as supercharges) have a structure
consistent with 12 dimensions, with a signature of (10,2). The extra 2
dimensions beyond 10 are hidden from the point of view of string theory. One
of them is spacelike and the other is timelike.

\item  As a consequence of central extensions of the superalgebra, $p$%
-branes naturally become part of the fundamental theory, and their
interaction with $p+1$ forms in supergravity are deduced. These $p$-branes
contribute to the non-perturbative multiplets demanded by $U$-duality and
hidden higher dimensions.

\item  The structure of (broken) symmetries associated with hidden
dimensions or $U$-duality, the groups, and a classification scheme for the
non-perturbative states emerge naturally from the structure of the
superalgebra. The relations between the hidden symmetries may be described
schematically as follows 
\begin{equation}
\begin{array}{l}
\mathrel{\mathop{\stackunder{c\ { }compact\,+\,2\, { }hidden\, {
}dims.}{SO(c+1,1)}}\limits_{\downarrow }}\otimes \mathrel{\mathop{SO(d-1,1)}
\limits_{spacetime}}\,\,\rightarrow \mathrel{\mathop{\left. \begin{array}{c}
{multiplets\, of} \\ {charges,\, states} \end{array} \right.
}\limits_{\uparrow }} \\ 
\left. 
\begin{array}{l}
\left. 
\begin{array}{l}
SO(c+1)_{\,\,\,1\,hidden\,dim} \\ 
SO(c)_L\otimes SO(c)_R
\end{array}
\right\} \,\,\,\,\,\,\rightarrow \,\,\,\,\,\,\,\mathrel{\mathop{%
\stackrel{maximal\,compact}{K}}\limits_{(duality)}} \\ 
\,\,\,\,\,\,\,\,\,\,\,\,\,\,\,\,\,\,\stackrel{\uparrow }{\mathrel{%
\mathop{SO(c,c)}\limits_{(T-duality)}}}
\end{array}
\right\} \rightarrow  \mathrel{\mathop{U}\limits_{(duality)}}
\end{array}
\label{subgroups}
\end{equation}

\item  Furthermore, one may start with perturbative string states, but then
add non-perturbative states that are needed in order to provide a basis for
the underlying superalgebra and its isometries $K$ and $SO(c+1,1)$. This is
a method of classifying the a-priori unknown non-perturbative states.
\end{enumerate}

It has been known since the 1970's that the structure of the superalgebra of
type IIA in 10 dimensions is intimately connected to 11 dimensions. In the
context of string duality this led to a possible ``M-theory'' with signature
(10,1) \cite{townsend11}-\cite{horavawitten}. Actually, there seems to be
room for (10,2) according to the general properties of the superalgebra
discussed below and other more specific arguments given elsewhere\footnote{%
The possibility of (10,2) emerged sometime ago \cite{duffbs}. In connection
with duality, it was discussed in a conference talk \cite{ibjapan}, where
the (10,2) behavior of the central extensions and the relation to duality
was emphasized. In more recent developments \cite{vafaf} \cite{martinecf}
other aspects of (10,2) in more detailed theories, such as ``F-theory'' have
been discussed.}, and the fact that type IIB in (9,1) can be related to type
IIA in (9,1) by a T-duality transformation in (10,2).

The points above will be the main topics of this paper. The last point
is supported by previous work \cite{ib11d},\cite{ibyank},\cite
{sendb}, in which consistency between non-perturbative U-multiplets and 11D
(broken) multiplets was analysed. In this paper we point out the possibility
12D (broken) multiplets.

\section{32 supercharges and (10,2)}

It is well known that the maximum number of supercharges in a physical
theory is 32. This constraint is obtained in four dimensions by requiring
that supermultiplets of massless particles should not contain spins that
exceed 2. Assuming that the four dimensional theory is related to a higher
dimensional one, then the higher theory can have at most 32 real
supercharges. Denote the 32 supercharges by $Q_\alpha ^a$, where $%
a=1,2,\cdots N$, and $\alpha $ is the spinor index in $d$-dimensions. For
example, in $d=11$ there is a single 32-component Majorana spinor (N=1), in
D=10 there are two 16-component Majorana-Weyl spinors (N=2), etc. down to
D=4 where there are eight 4-component Majorana spinors (N=8). It is
important to note that 32 corresponds to counting {\it real} components of
spinors.

In 12 dimensions the Weyl spinor also has 32 components since $\frac
122^{12/2}=32,$ but when the signature is $\left( 11,1\right) $ the spinor
is complex and has 64 real components. Therefore, as long as we consider a
single time coordinate, $d=11$ is the highest allowed dimension. However, if
the signature is $\left( 10,2\right) ,$ it is possible to impose a Majorana
condition that permits a real 32-component spinor \footnote{%
A quick way to see this is to use Bott periodicity to relate the properties
of the spinors with signatures (2,2)$\sim \left( 10,2\right) $. For $%
SO\left( 2,2\right) $ the Weyl spinor is real since $SO\left( 2,2\right)
\sim SL(2,R)\times SL(2,R).$ Hence it is also real for signature $\left(
10,2\right) .$ Another quick remark is that the Lorentz group $SO(n,1)$ and
the conformal group $SO(n,2)$ for $n$ spacelike dimensions have the same
same spinor representations. Hence the 32 dimensional spinor is a basis for
both $SO(10,1)$ and $SO(10,2).$ Since it is real for 11D it must also be
real for 12D with signature $(10,2)$.}. Thus, a price to pay to go beyond 11
dimensions is to consider a second time-like coordinate. It is not clear
that traditional unphysical problems of 2 time coordinates may not be
circumvented in some unknown, sufficiently constrained, theory. Hence we may
entertain the possibility of $\left( 10,2\right) $ if there are some
benefits for doing so, provided physical inconsistencies are eliminated.
Beyond 12 dimensions the spinor is too large, and therefore we cannot
consider $d>12$.

We need to discuss the theory and analyze its content of hidden dimensions.
For example, type IIA string theory with signature (9,1) will appear to be a
toroidal compactification from (10,2) on $R^{9,1}\otimes T^{1,1}$ where the
extra dimensions with signature $\left( 1,1\right) $ are both considered
hidden, one of them spacelike and the other timelike. More generally we will
consider toroidal compactifications on $R^{d-1,1}\otimes T^{c+1,1}$ where $d$
is the number of ordinary Minkowski spacetime dimensions and $c$ is the
number of compactified {\it string} dimensions, while the two hidden
dimensions are counted as extra, so that $d+c+2=12.$ The 32 spinors $%
Q_\alpha ^a$ may then be classified as the spinor for $SO(d-1,1)\otimes
SO(c+1,1).$ The index $a$ corresponds to the spinor of $SO(c+1,1).$ This
group is not necessarily a symmetry, but it helps to keep track of the
compactified dimensions, including the hidden ones. Furthermore, the same $a$
index will be reclassified later under the maximal compact subgroup $K$ of $%
U $-duality, thus providing a bridge between duality and higher hidden
dimensions. The supercharges labelled in this way are listed in Table I in
various dimensions (at this stage of the discussion the $K$ content of Table
I should be ignored). The fact that the same index $a$ is classified in {\it %
irreducible} representations of the hidden symmetries of two types is a
significant point for the arguments in the rest of the paper.

Consider the maximally extended algebra of the 32 supercharges in various
dimensions in the form 
\begin{equation}
\left\{ Q_\alpha ^a,Q_\beta ^b\right\} =\delta ^{ab}\gamma _{\alpha \beta
}^\mu \,\,P_\mu +\sum_{p=0,1,\cdots }\gamma _{\alpha \beta }^{\mu _1\cdots
\mu _p}\,\,\,Z_{\mu _1\cdots \mu _p}^{ab}.  \label{supergeneral}
\end{equation}
Since the left side is the symmetric product of 32 supercharges, the right
side can have at most $\frac 1232\times 33=528$ independent generators. The
indices $ab$ on $Z_{\mu _1\cdots \mu _p}^{ab}$ are either symmetrized or
antisymmetrized and have the same permutation symmetry as $\alpha \beta $ in 
$\gamma _{\alpha \beta }^{\mu _1\cdots \mu _p}.$ The central extensions $%
Z_{\mu _1\cdots \mu _p}^{ab}\,$ are assumed \footnote{%
For simplicity we assume commuting central extensions. There are more
involved versions of the extended superalgebra in which some of the central
extensions do not commute with $Q_\alpha ^a,$ or with each other, etc. \cite
{sezgin}. We might expect that the non-commuting cases may arise for curved
backgrounds and non-toroidal compactifications that are not discussed here.}
to commute with $Q_\alpha ^a,P_\mu ,$ but they are tensors of the Lorentz
group and hence do not commute with it. According to a theorem of Haag et.
al. \cite{haag}, there can be only Lorentz scalar central charges in a
unitary theory in four dimensions, for interactions of {\it point-like
particles (}$p=0${\it )}. However, as will become clear below, in the
presence of $p$-branes new interactions that permit Lorentz tensors $Z_{\mu
_1\cdots \mu _p}^{ab}$ are present in theories with a unitary S-matrix (e.g.
string theory), indicating that the theorem \cite{haag} does not apply to
extended objects. 

\[
\begin{tabular}{|c|c|c|c|c|c|c|c|c|}
\hline
$^{\frac{c+1,1}{d-1,1}}$ & $_{\Sp SO(c+1,1)  \\ ({or\, \thinspace }%
K\,)\,\,\otimes \,\,  \\ SO(d-1,1)  \endSp }^{32\,\,Q_\alpha ^a}$ & $\Sb %
P^m,Z^{mn}  \\ X^{mnlqr}  \endSb \Sp p=0  \\ .  \endSp $ & $\Sb P_\mu ,Z_\mu
^n  \\ X_\mu ^{nlqr}  \endSb \Sp p=1  \\ .  \endSp $ & $\Sb Z_{\mu v}  \\ %
X_{\mu \nu }^{lqr}  \endSb \Sp p=2  \\ .  \endSp $ & $\Sb .  \\ X_{\mu \nu
\lambda }^{qr}  \endSb \Sp p=3  \\ .  \endSp $ & $\Sb .  \\ X_{_{\mu _1.\mu
_4}}^r  \endSb \Sp p=4  \\ .  \endSp $ & $\Sb .  \\ X_{\mu _1.\mu _5} 
\endSb \Sp p=5  \\ .  \endSp $ & $\frac{^U}K$ \\ \hline
$\Sp A  \\ \frac{1,1}{9,1}  \endSp $ & $^{\left( {\bf \pm ,16}\right) }$ & $
\Sp 1+0  \\ +0  \endSp $ & $\Sp 1+1  \\ +0  \endSp $ & $\Sp 1  \\ +0  \endSp 
$ & $^0$ & $^1$ & $\Sp 1^{+}  \\ +1^{-}  \endSp $ & $\frac{^{SO(1,1)}}{Z_2}$
\\ \hline
$\Sp B  \\ \frac{1,1}{9,1}  \endSp $ & $^{\left( _{+}^{+}{\bf ,16}\right) }$
& $\Sp 0+0  \\ +0  \endSp $ & $\Sp 1+2  \\ +0  \endSp $ & $\Sp 0  \\ +0 
\endSp $ & $^1$ & $^0$ & $\Sp 1^{+}  \\ +2^{+}  \endSp $ & $\frac{^{SL(2,R)}%
}{^{SO(2)}}$ \\ \hline
$^{\frac{2,1}{8,1}}$ & $^{\left( {\bf 2,16}\right) }$ & $\Sp 2+1  \\ +0  \\ =%
{\bf 3}  \\ \approx {\bf 2+1}  \endSp $ & $\Sp 1+2  \\ +0  \\ ={\bf 3}  \\ %
\approx {\bf 2+1}  \endSp $ & $\Sp 1+0  \\ ={\bf 1}  \\ \approx {\bf 1} 
\endSp $ & $^1$ & $\Sp \left[ 1\right]  \\ +2  \\ ={\bf 3}  \\ \approx {\bf 2%
}  \\ {\bf +1}  \endSp $ & $\Sp \left( 1\right)  \\ _{move}  \endSp $ & $%
\frac{\Sp SL(2)\otimes  \\ SO(1,1)  \endSp }{\Sp SO(2)  \\ \otimes Z_2 
\endSp }$ \\ \hline
$^{\frac{3,1}{7,1}}$ & $^{\Sp \left( \left( {\bf 2},0\right) {\bf ,8}%
^{+}\right)  \\ \left( \left( 0,{\bf 2}\right) {\bf ,8}^{-}\right)  \endSp }$
& $\Sp 3+3  \\ +0  \\ ={\bf 6}  \\ \approx {\bf 3}^{{\bf +}}  \\ +{\bf 3}%
^{-}  \endSp $ & $\Sp 1+3  \\ +0  \\ =\left( {\bf 2,2}\right)  \\ \approx 
{\bf 3+1}  \endSp $ & $\Sp 1+1  \\ ={\bf 1+1}  \\ \approx {\bf 1+1}  \endSp $
& $\Sp 3+\left[ 1\right]  \\ =\left( {\bf 2,2}\right)  \\ \approx {\bf 3+1} 
\endSp $ & $\Sp 3^{+}  \\ +3^{-}  \\ ={\bf 6}  \\ \approx {\bf 3}^{+}  \\ +%
{\bf 3}^{-}  \endSp $ & $\Sp \left( 1\right)  \\ _{move}  \endSp $ & $^{%
\frac{\Sp SL(3)  \\ \otimes SL(2)  \endSp }{\Sp SO(3)  \\ \otimes U(1) 
\endSp }}$ \\ \hline
$^{\frac{4,1}{6,1}}$ & $^{\left( {\bf 4,8}\right) }$ & $\Sp 4+6  \\ +0  \\ =%
{\bf 10}  \\ \approx {\bf 10}  \endSp $ & $\Sp 1+4  \\ +1  \\ ={\bf 5+1}  \\ %
\approx {\bf 5+1}  \endSp $ & $\Sp 1+4  \\ +\left[ 1\right]  \\ ={\bf 5+1} 
\\ \approx {\bf 5}+{\bf 1}  \endSp $ & $\Sp 6  \\ +\left[ 4\right]  \\ ={\bf %
10}  \\ \approx {\bf 10}  \endSp $ & $\Sp \left( 4\right)  \\ _{move} 
\endSp $ & $\Sp \left( 1\right)  \\ _{move}  \endSp $ & $\frac{^{SL(5)}}{%
^{SO(5)}}$ \\ \hline
$_{\frac{5,1}{5,1}}$ & $\Sb \left( {\bf 4,4}^{*}\right)  \\ \left( {\bf 4}%
^{*},{\bf 4}\right)  \endSb $ & $_{\Sp 5+10  \\ +1  \\ ={\bf 1+15}  \\ %
\approx \left( {\bf 4,4}\right)  \endSp }$ & $_{\Sp  \\ 1+5+  \\ 5+\left[
1\right]  \\ =2\times {\bf 6}  \\ \approx \left( {\bf 0,5}\right)  \\ %
+\left( {\bf 5,0}\right)  \\ +2\left( {\bf 0,0}\right)  \endSp }$ & $_{\Sp %
1+10  \\ +\left[ 5\right]  \\ ={\bf 1+15}  \\ =\left( {\bf 4,4}\right) 
\endSp }$ & $_{\Sp 10^{+}  \\ +10^{-}  \\ ={\bf 10}^{+}  \\ +{\bf 10}^{-} 
\\ \approx ({\bf 10,}1{\bf )}  \\ +(1,{\bf 10})  \endSp }$ & $_{\Sp \left(
5\right)  \\ _{move}  \endSp }$ & $_{\Sp \left( 1\right)  \\ _{move}  \endSp %
}$ & $_{\frac{SO(5,5)}{\Sp SO(5)  \\ \otimes SO(5)  \endSp }}$ \\ \hline
$^{\frac{6,1}{4,1}}$ & $^{\left( {\bf 8,4}\right) }$ & $\Sp 6+15  \\ +6  \\ %
+\left[ 1\right]  \\ ={\bf 7+21}  \\ \approx {\bf 27}+{\bf 1}  \endSp $ & $
\Sp 1+6  \\ +15+\left[ 6\right]  \\ ={\bf 7+21}  \\ \approx {\bf 27+1} 
\endSp $ & $\Sp 1+20  \\ +\left[ 15\right]  \\ ={\bf 1+35}  \\ \approx {\bf %
36}  \endSp $ & $\Sp \left( 15\right)  \\ _{move}  \endSp $ & $\Sp \left(
6\right)  \\ _{move}  \endSp $ & $\Sp \left( 1\right)  \\ _{move}  \endSp $
& $^{\frac{^{E_{6(6)}}}{^{USp(8)}}}$ \\ \hline
$^{\frac{7,1}{3,1}}$ & $^{\Sp ({\bf 8}^{+}{\bf ,(2,0))}  \\ ({\bf 8}^{-}{\bf %
,}(0,{\bf 2)})  \endSp }$ & $\Sp 7+21  \\ +21  \\ +\left[ 7\right]  \\ {\bf %
=28+28}  \\ \approx {\bf 28}_c  \endSp $ & $\Sp 1+7  \\ +35  \\ +\left[
21\right]  \\ ={\bf 8+56}  \\ \approx {\bf 63+1}  \endSp $ & $\Sp 1^{\pm } 
\\ +35^{\pm }  \\ ={\bf 1}^{\pm }  \\ +{\bf 35}^{\pm }  \\ \approx {\bf 36}%
_c  \endSp $ & $\Sp \left( 21\right)  \\ _{move}  \endSp $ & $\Sp \left(
7\right)  \\ _{move}  \endSp $ & $^0$ & $\frac{^{E_{7(7)}}}{^{SU(8)}}$ \\ 
\hline
$^{\frac{8,1}{2,1}}$ & $^{\left( {\bf 16,2}\right) }$ & $\Sp 8+28  \\ +56 
\\ +\left[ 28\right]  \\ ={\bf 36+84}  \\ \approx {\bf 120}  \endSp $ & $\Sp %
1+8+70  \\ +\left[ 1+56\right]  \\ ={\bf 1+9}  \\ {\bf +126}  \\ \approx 
{\bf 135}  \\ {\bf +1}  \endSp $ & $\Sp \left( 1+56\right)  \\ _{move} 
\endSp $ & $\Sp \left( 28\right)  \\ _{move}  \endSp $ & $^0$ & $^0$ & $%
\frac{^{E_{8(8)}}}{^{SO(16)}}$ \\ \hline
\end{tabular}
\]
\[
\rm{Table I.\, Classification\, of\, }Q_\alpha ^a\rm{\, and\, Z}_{\mu _1\cdots \mu
_p}^{ab}\rm{\, under\, 11D\, (or\, 12D)\, and\, }K\rm{.} 
\]
In (10,2) dimensions we will use $M=0^{\prime },0,1,2,\cdots ,10$ for the
space index instead of $\mu .$ In the 32$\times 32$ representation
(equivalent to {\it chirally projected} 64$\times 64$) only the 2- and 6-
index gamma matrices $\gamma _{\alpha \beta }^{M_1M_2}$ and $\gamma _{\alpha
\beta }^{M_1\cdots M_6}$ are symmetric in $\alpha \beta ,$ and furthermore $%
\gamma _{\alpha \beta }^{M_1\cdots M_6}$ is self dual (one gamma matrix
index has been lowered by multiplying with the charge conjugation matrix).
The remaining $\gamma _{\alpha \beta }^{M_1\cdots M_p}$ do not have definite
symmetry or antisymmetry in $\alpha \beta .$ Therefore, in 12 dimensions, on
the right hand side of (\ref{supergeneral}) there can be no $P_M,$ and the
528 generators consist of the antisymmetric tensors $Z_{M_1M_2}$ and $%
Z_{M_1\cdots M_6}^{+}$ which is self dual$.$ The number of components in
each is 
\begin{equation}
\begin{array}{l}
\frac{12\times 11}2=66,\quad \frac 12\frac{12\times 11\times 10\times
9\times 8\times 7}{1\times 2\times 3\times 4\times 5\times 6}=462,
\end{array}
\end{equation}
respectively. Upon compactification to (10,1) we rewrite the 12D index $%
M=(0^{\prime },\mu )$ where $\mu =0,1,2,\cdots ,10$ is an 11D index. Then we
have (suppressing the $0^{\prime }$ index) 
\begin{eqnarray}
Z_{M_1M_2} &\rightarrow &P_\mu \oplus Z_{\mu _1\mu _2}\quad 66=11+55 \\
Z_{M_1\cdots M_6}^{+} &\rightarrow &X_{\mu _1\cdots \mu _5}\quad \quad \quad
462=462  \nonumber
\end{eqnarray}
which are the momenta and central charges in 11 dimensions pointed out in 
\cite{townsend}.

Continuing the compactification process to lower dimensions on $%
R^{d-1,1}\otimes T^{c+1,1},$ each eleven dimensional index $\mu $ decomposes
into $\mu \oplus m$ where $\mu $ is in $d$ dimensions and $m$ is in $%
c+1=11-d $ dimensions. Then each 11 dimensional tensor decomposes as follows 
\begin{eqnarray}
P_\mu &\rightarrow &P_\mu \oplus P_m \ \ \ \ \ \ \ 11=d+(c+1) \nonumber \\
Z_{\mu \nu } &\rightarrow &Z_{\mu \nu }\oplus Z_\mu ^n\oplus Z^{mn} \\
X_{\mu _1\cdots \mu _5} &\rightarrow &X_{\mu _1\cdots \mu _5}\oplus X_{\mu
_1\cdots \mu _4}^{m_1}\oplus X_{\mu _1\mu _2\mu _3}^{m_1m_2}  \nonumber \\
&&\oplus X_{\mu _1\mu _2}^{m_1m_2m_3}\oplus X_{\mu _1}^{m_1\cdots m_4}\oplus
X^{m_1\cdots m_5}.  \nonumber
\end{eqnarray}
For example for $(d=10, \ \ c=0)$ 
the type IIA superalgebra is recovered, with the 528
operators ($P_\mu ,P_{10},Z_{\mu \nu },Z_\mu ,X_{\mu _1\cdots \mu _4},X_{\mu
_1\cdots \mu _5}^{\pm })$ where the $\pm $ indicate self/antiself dual
respectively. In Table I in each row labelled by $(d-1,1)/(c+1,1)$ the
numbers of each central extension of $P,Z,X$ type with $p$ Lorentz indices
is indicated (these are the numbers that are not in bold characters). Since
each of the $P,Z,X$ are antisymmetric tensors in $c+1$ dimensions these
numbers correspond to representations of $SO(c+1)$ (which includes rotations
into one of the extra dimensions). As we go to lower dimensions one must use
the duality between $p$ indices and $d-p$ indices to reclassify and count
the central extensions 
$Z_{\mu _1\cdots \mu _p}^{ab}\sim Z_{\mu _1\cdots \mu _{d-p}}^{ab}$. 
In the table a number
in parenthesis means that it should be omitted from there and instead moved
in the same row to the location where the same number appears in brackets.
This corresponds to the equivalence of $p$ indices and $d-p$ indices. When $%
p=d-p$ there are self-dual or anti-self-dual tensors. Their numbers are
indicated with additional superscripts $\pm $ in the form $1^{\pm },2^{\pm
},3^{\pm },10^{\pm },35^{\pm }$ wherever they occur.

The total number of central extensions $P,Z,X$ found according to this
compactification procedure for each value of $p$ are indicated in Table-I in
bold characters. These totals are the same numbers found by counting the 
{\it number} of possibilities $ab$ on $Z_{\mu _1\cdots \mu _p}^{ab}.$ The
bold numbers following the $=$ sign correspond to representations of $%
SO(c+1,1)$ (making a connection to 12D) and those following the $\approx $
sign correspond to representations of $K$ (to be discussed later in
connection to duality).

\section{Central charges and p-branes}

What is the meaning of the $p$-form central extension $Z_{\mu _1\cdots \mu
_p}^{ab}$? Since this is a charge in a global algebra, there ought to exist
a ($p+1$)-form local current $J_{\mu _0\mu _1\cdots \mu _p}^{ab}\left(
x\right) $ whose integral over a space-like surface embedded in $d$%
-dimensions gives 
\begin{equation}
Z_{\mu _1\cdots \mu _p}^{ab}=\int d^{d-1}\Sigma ^{\mu _0}\,\,J_{\mu _0\mu
_1\cdots \mu _p}^{ab}\left( x\right) .
\end{equation}
The current couples to the fields of low energy physics (i.e. supergravity).
In the case of usual central charges that are Lorentz singlets $Z^{ab}$
(i.e. $p=0)\,$the current is associated with charged particles. Such a
current may be constructed as usual from worldlines (or equivalently from
local fields) as follows 
\begin{equation}
J_{\mu_0} ^{ab}\left( x\right) =\int d\tau \sum_iz_i^{ab}\,\delta ^d\left(
x-X^i\left( \tau \right) \right) \,\,\partial _\tau X_{\mu_0} ^i\left( \tau
\right) .
\end{equation}
The $z_i^{ab}$ are the charges of the particles labelled by $i.$ This
current couples in the action 
to a gauge field $A_{ab}^{\mu_0} $, and it appears
as the source in the equation of motion of the abelian\footnote{%
The gauge fields are Abelian since we assumed commuting central charges. As
noted in a previous footnote, a non-Abelian version is expected if the
background is curved rather than flat.} gauge field 
\begin{eqnarray}
\begin{array}{l}
S\sim \sum_i\int d\tau \,A_{ab}^\mu \left( X^i\left( \tau \right) \right)
\,\partial _\tau X_\mu ^i\left( \tau \right) \,\,z_i^{ab} \\ 
\,\,\,\,=\int d^dx\,\,A_{ab}^\mu \left( x\right) \,J_\mu ^{ab}\left( x\right)
\\ 
\partial _{\lambda \,\,}\partial ^{[\lambda }A_{ab}^{\mu ]}\left( x\right)
=J_{ab}^\mu \left( x\right) .
\end{array}
\end{eqnarray}
Therefore, there are as many gauge fields as there are central extensions of
type $p=0.$ These gauge fields occur as massless particles in the
NS-NS and R-R sectors of the superstring. The charges $Z^{ab}$associated with the
NS-NS sector occur perturbatively in string theory (Kaluza-Klein momenta and
winding numbers), but the charges associated with the R-R sector are
non-perturbative from the point of view of string theory (topological
solitonic charges). On the other hand, from the point of view of the
superalgebra they occur at an equal footing, and will be treated on an equal
basis from the point of view of the (broken) symmetries that we discuss
later.

Central charges with $p\geq 1$ have been usually omitted in past discussions
due to the theorem in \cite{haag}. The theorem allows only $p=0$ central
extensions. This was derived under the assumption of a unitary S-matrix
based on point-like interactions in 4-dimensions. However, let us now
discuss the implications of central charges in the presence of extended
objects and in any dimension. For $p=1,$ the central extension is a vector 
$Z_{\mu_1} ^{ab}$, which requires a local 
current that is an antisymmetric tensor 
$J_{{\mu_0} {\mu_1} }^{ab}\left( x\right) $ in the Lorentz indices\footnote{%
New symmetric tensors, other than the symmetric stress tensor $\delta
^{ab}T_{{\mu_0} {\mu_1} }(x)$ associated with the momentum 
$\delta ^{ab}P_{\mu_1} $ ($\sim Z_{\mu_1} ^{ab}$), , are
not allowed in the superalgebra, since they would couple to new
``gravitons''.}. An antisymmetric current cannot be constructed from
particles but it can be constructed from{\it \ strings }as follows 
\begin{equation}
J_{{\mu_0} {\mu_1} }^{ab}\left( x\right) 
=\int d\tau d\sigma \sum_iz_i^{ab}\,\delta
^d\left( x_\mu-X^i_\mu\left( \tau ,\sigma \right) \right) 
\,\,\partial _\tau X_{[{\mu_0}
}^i\left( \tau ,\sigma \right) 
\,\partial _\sigma X_{{\mu_1} ]}^i\left( \tau
,\sigma \right) \,\,.
\end{equation}
$z_i^{ab}$ is the charge of the $i^{th}$ string$.$ Just like the particles
discussed above, the charged strings also are expected to form a multiplet
of the (broken) symmetries, and interact with the low energy supergravity
fields through antisymmetric gauge potentials 
$B_{ab}^{{\mu_0} {\mu_1} }\left(
x\right) ,$ with an action 
\begin{eqnarray}
S &\sim &\sum_i\int d\tau d\sigma \,B_{ab}^{\nu \mu }\left( X^i\left( \tau
,\sigma \right) \right) \,\partial _\tau X_{[\nu }^i\left( \tau ,\sigma
\right) \,\partial _\sigma X_{\mu ]}^i\left( \tau ,\sigma \right) \,z_i^{ab}
\\
&=&\int d^dx\,\,B_{ab}^{\nu \mu }\left( x\right) \,J_{\nu \mu }^{ab}\left(
x\right)  \nonumber
\end{eqnarray}
In this expression one can recognize the familiar string coupling to an
antisymmetric tensor in the world sheet formulation. The equation of motion
for $B_{ab}^{\nu \mu }\left( x\right) $ involves the (abelian) gauge
invariant field strength $H_{ab}^{\lambda \nu \mu }=\partial ^{[\lambda
}B_{ab}^{\nu \mu ]}\left( x\right) $ and the above current as a source 
\begin{equation}
\partial _{\lambda \,\,}\partial ^{[\lambda }B_{ab}^{\nu \mu ]}\left(
x\right) =J_{ab}^{\nu \mu }\left( x\right) .
\end{equation}
A well known example is type IIB superstring with its two antisymmetric
tensors. In this case the $ab$ indices on $B_{ab}^{\nu \mu }\left( x\right) $
correspond to a symmetric traceless 2$\times 2$ matrix. More antisymmetric
tensors are found in compactifications to lower dimensional string theories.

This example also shows that central extensions that are not Lorentz
singlets are present in a unitary theory with non-trivial scattering.
Therefore the theorem in \cite{haag}, while valid for point particle
interactions, should not be applicable in the presence of $p$-branes and
their interactions.

The generalization to the higher values of $p$ is straightforward$:$ In
order to have a charge that is a $p$-form we need a current $J_{\mu _0\mu
_1\cdots \mu _p}^{ab}\left( x\right) $ that is a $(p+1)$-form$.$ This in
turn requires a $p$-brane to construct the current, 
\begin{eqnarray}
J_{\mu _0\mu _1\cdots \mu _p}^{ab}\left( x\right) &=&\int d\tau d\sigma
_{1\cdots }d\sigma _p\sum_iz_i^{ab}\,\delta ^d\left( x-X^i(\tau ,\sigma
_{1,}\cdots \sigma _p)\right) \,\, \\
&&\times \partial _\tau X_{[\mu _0}^i\cdots \,\partial _{\sigma _p}X_{\mu
_p]}^i(\tau ,\sigma _{1,}\cdots \sigma _p)\,\,,  \nonumber
\end{eqnarray}
and its coupling to supergravity fields requires a $(p+1)$-form gauge
potential $A^{ab}_{\mu _0\mu _1\cdots \mu _p}\left( x\right) $ such that 
\begin{eqnarray}
S &\sim &\int d^dx\,\,A_{ab}^{\mu _0\mu _1\cdots \mu _p}\left( x\right)
\,J_{\mu _0\mu _1\cdots \mu _p}^{ab}\left( x\right) \\
&=&\sum_i\int d\tau d\sigma _{1\cdots }d\sigma _p\,A_{ab}^{\mu _0\mu
_1\cdots \mu _p}\left( X^i\right) \,\,\partial _\tau X_{[\mu _0}^i\cdots
\,\partial _{\sigma _p}X_{\mu _p]}^i\,\,z_i^{ab},  \nonumber
\end{eqnarray}
and 
\begin{equation}
\partial _{\lambda \,\,}\partial ^{[\lambda }A_{ab}^{\mu _0\mu _1\cdots \mu
_p}\left( x\right) =J_{ab}^{\mu _0\mu _1\cdots \mu _p}\left( x\right) .
\end{equation}
As is well known by now there are perturbative as well as non-perturbative
couplings of $p$-branes to supergravity in various dimensions. Hence the $%
Z_{\mu _1\cdots \mu _p}^{ab}$ are present in the superalgebra and they
correspond simply to the charges of $p$-branes. The classification of their $%
ab$ indices under duality groups is the subject of the next section, but
here we already see that there is a one to one correspondence between the $p$%
-forms $Z_{\mu _1\cdots \mu _p}^{ab}$ and the ($p+1)$-form gauge potentials $%
A_{ab}^{\mu _0\mu _1\cdots \mu _p}$ that appear as massless states in string
theory in the NS-NS or R-R sectors.

The main message is that from the point of view of the superalgebra all $p$%
-branes appear to be at an equal footing. Isometries of the superalgebra
that will be discussed below treat them equally and may mix them with each
other in various compactifications. The theory in $d$ dimensions has $(p+1)$%
-forms $A_{ab}^{\mu _0\mu _1\cdots \mu _p}$ which appear as massless
particles in the string version of the fundamental theory. These act as
gauge potentials and couple at low energies to charged $p$-branes. This
generates a non-trivial central extension $Z_{\mu _1\cdots \mu _p}^{ab}$ in
the superalgebra. The number of such central extensions ($ab$ indices) is in
one-to-one correspondence with the number of the $(p+1)$-forms $A_{ab}^{\mu
_0\mu _1\cdots \mu _p}$, and these numbers can be obtained by counting the
possible combination of (symmetric/antisymmetric) indices $ab$ associated
with the supercharges.

\section{Reclassification and duality}

In the discussion above we concentrated on the 11D (or 12D) content of the
supercharges and the central extensions. We now turn to duality. In string
theory the $T$-duality group is directly related to the number of
compactified left/right string dimensions. In our notation the number of
compactified string dimensions is $c.$ Therefore, for a string of type II,
it is 
\begin{equation}
T=SO(c,c).
\end{equation}
Its maximal compact subgroup is 
\begin{equation}
k=SO(c)_L\otimes SO(c)_R.
\end{equation}
where $L,R$ denote left/right movers respectively
\footnote{
The notation for duality groups, such as $SO(c,c),$ is used somewhat loosely
in this paper, for brevity. The $T,U$ duality groups mentioned in this paper are
supposed to be interpreted as discrete groups, such as $SO(c,c,Z),$ etc.
This is not apparent from the superalgebra point of view, but is true in
string theory. Under T-duality transformations the 
quantized Kaluza-Klein and winding
numbers of string states transform into each other under $SO(c,c,Z)$. In addition
there is an induced transformation on the oscillators in the internal dimensions
under the subgroup $k=SO(c)_L\otimes SO(c)_R$, where the effective parameters of
the induced transformation depend on the discrete $SO(c,c,Z)$ as well as the 
torus parameters $G_{ij},B_{ij}$, and hence it is equivalent to being continuous. 
Therefore, because of
T-duality, all perturbative string states must fall into linear representations 
of $k=SO(c)_L\otimes SO(c)_R$, which is larger than the $SO(c)$ expected naively. 
In a similar sense, the transformations under $K$ are also equivalent to being
continuous, even though those of $U$ are discrete. 
For a clarification of these points see \cite{ibyank}. }. 
The supercharges $Q_\alpha ^a$ naturally know about this
group, since they too can be split into left/right movers in even $d$
dimensions: then the index $a$ on left/right chiral charges $Q_\alpha ^a$
corresponds precisely to the spinor index of $SO(c)_L\otimes SO(c)_R.$ For
odd $d$ dimensions the same is true, but the L/R split is defined by going
to the next smaller value of $c.$

For example, in four dimensions the $N=8$ real Majorana spinors are
rewritten as $8$ pseudo-real Weyl spinors of left or right type that are
each other's complex conjugates. In Table I these were classified as
pseudo-real representations 
\begin{equation}
\begin{array}{l}
(8^{+},(2,0)),\,\,(8^{-},(0,2)) \\ 
SO(7,1)_{hidden}\otimes SO(3,1)_{space},
\end{array}
\end{equation}
Now we reclassify them as 
\begin{equation}
\begin{array}{l}
([(4,0)+(0,4^{*})],(2,0)),\,\,([(4^{*},0)+(0,4)],(0,2)) \\ 
\left( SO(6)_L\times SO(6)_R\right) _{k\subset T}\otimes SO(3,1)_{space}.
\end{array}
\end{equation}
The common internal group in $SO(7,1)_{hidden}$ and $SO(6)_L\times SO(6)_R$
is $SO(6)$, but besides this common subgroup these two groups are not
related to each other by group/subgroup relationships. Thus their
transformations on the physical states of the theory must act on rather
different modules that have intersections with each other.

More generally for any dimension, investigating the supercharges listed in
Table I shows that, the index $a$ that was classified there under the hidden
non-compact group $SO(c+1,1)_{hidden}$ can be reclassified under the
perturbatively explicit maximal compact subgroup $k\subset T$ of T-duality, $%
k=SO(c)_L\otimes SO(c)_R$ (see Table III in ref.\cite{ibyank})$.$ These two
groups are not subgroups of each other, but they do have a common subgroup $%
SO(c).$ Recall that $c$ is the number of compactified string dimensions
(other than the two hidden dimensions), and $SO(c)$ is the (broken) rotation
group in these internal dimensions.

In each case one may notice that the $N$ supercharges $Q_\alpha ^a$
transform irreducibly under $SO(c+1,1)_{hidden}$, but reducibly under $%
k=SO(c)_L\times SO(c)_R$. However, we can obtain an irreducible
representation $Q_\alpha ^a$ by defining a larger {\it compact} group $K$
that contains $k$, as well as the maximal compact part of $%
SO(c+1,1)_{hidden} $. That is 
\begin{equation}
K\supset SO(c)_L\otimes SO(c)_R\rm{ \thinspace \thinspace and\thinspace
\thinspace \thinspace }K\supset SO(c+1).
\end{equation}
Thus we look for the {\it compact} group $K$ that contains $SO(c)_L\times
SO(c)_R,\,$ $SO(c+1)$ and that has an{\it \ irreducible} representation for
the index $a$ (total dimension $N$)$.$ Note that the group $K$ must mix one
extra dimension with others. Furthermore, the central extensions of type $%
P,Z,X$ that already display the extra dimension have to fall into
representations of $K$ that contain them. The {\it minimal} compact $K$ that
we find through this reasoning is listed in the last column of
Table I. By virtue of containing $k\subset T$ the group $K\supset k$ must be
related to a larger group of duality $U$ that contains $T$. After finding $K$
as described, $U$ is determined uniquely by looking for the smallest
non-compact group that contains $SO(c,c)$ and for which $K$ is the maximal
compact subgroup. The subgroup hierarchy that emerges is given in eq.(\ref
{subgroups}). For example in four dimensions (with $d=4,\,\,c=6)$ it is 
\begin{equation}
\begin{array}{l}
\mathrel{\mathop{SO(7,1)_{hidden}}\limits_{\downarrow }}\otimes
SO(3,1)\rightarrow \,a:\left. 
\begin{array}{c}
(8^{+},(2,0)) \\ 
+(8^{-},(0,2))
\end{array}
\right. \\ 
\left. 
\begin{array}{l}
\left. 
\begin{array}{l}
SO(7)_{hidden} \\ 
SO(6)_L\otimes SO(6)_R
\end{array}
\right\} \rightarrow \stackrel{\uparrow }{SU(8)} \\ 
\stackrel{\uparrow }{T=SO(6,6)}
\end{array}
\right\} \rightarrow E_{7(7)}
\end{array}
\end{equation}
The $a$ index which was classified as the spinors $8^{\pm }$ under $SO(7,1)$
or as $\left[ (4,0)+(0,4^{*})\right] $ or $\left[ (4^{*},0)+(0,4)\right] $
under $k=SO(6)_L\otimes SO(6)_R$ is now reclassified as the $8$ or $8^{*}$
of $SU(8).$ This group is the minimal compact group containing both $SO(7)$
and $SO(6)_L\otimes SO(6)_R=SU(4)_L\otimes SU(4)_R.$ Furthermore, the
smallest non-compact group containing both $SU(8)$ and $SO(6,6)$ is $%
E_{7(7)}.\,$ This way of describing $K$ or $U$ does not use the details of
supergravity or string theory. It merely hinges on the number of
supercharges and their reclassifications in maximally irreducible
representations as described above. We emphasize that the scheme takes
advantage of the hidden dimensions.

Since the same $N$ dimensional basis of supercharges labelled by $a$ knows
about both duality and the hidden dimensions, this must provide a bridge for
relating properties of the states of the theory under both qualities. The
first consequence of this is the reclassification of the central extensions $%
Z_{\mu _1\cdots \mu _p}^{ab}.$ Previously they were classified under 11D\
(or 12D) as in Table I (the numbers following the = sign). 
But now the combination $ab$ corresponds to the
symmetric or antisymmetric product of the $N$ dimensional representation of $%
K.$ Therefore, {\it the central extensions are now also classified under} $%
\dot{K}$. The result is the total dimension listed in Table I 
(the numbers following the $\approx$ sign). These numbers
are indeed dimensions of irreducible multiplets under $K.$

For example in four dimensions the central extensions whose (real) numbers
are $56,63,72$ for $p=0,1,2$ respectively, are reclassified as the complex $%
{\bf 28}_c,$ real {\bf 63} and complex {\bf 36}$_c$ of $K=SU(8).$ These
correspond to the following combinations of the $SU(8)$ $ab$ indices on $%
Z_{\mu _1\cdots \mu _p}^{ab},$ recalling that $a\rightarrow 8$ or $8^{*}: $%
\begin{eqnarray}
p &=&0:\quad \left( 8\times 8\right) _{antisymm}=28_c  \nonumber
\label{Klist} \\
p &=&1:\quad \left( 8\times 8^{*}\right) =63+1  \label{klist} \\
p &=&2:\quad \left( 8\times 8\right) _{symmetric}=36_c  \nonumber
\end{eqnarray}
The $p=1$ singlet {\bf 1} corresponds to the momentum $P^\mu$.
The complex conjugates $28_c^{*},36_c^{*}$ contain the same real components
as $28_c,36_c$.
On the other hand these same total dimensions correspond to the irreducible
representations of $SO(7,1)_{hidden}$ as follows. Using the fact that the
supercharges can be viewed as the spinors $8^{+}\oplus 8^{-}$, their
products give the following $SO(7,1)_{hidden}$ representations for the
indices $ab$ on $Z_{\mu _1\cdots \mu _p}^{ab}$ 
\begin{eqnarray}
p &=&0:\quad \left( 8^{\pm }\times 8^{\pm }\right) _{antisymm}=28^{\pm } 
\nonumber \\
p &=&1:\quad \left( 8^{+}\times 8^{-}\right) =8_v+56_v
\label{olist} \\
p &=&2:\quad \left( 8^{\pm }\times 8^{\pm }\right) _{symmetric}=1^{\pm
}+35^{\pm }  \nonumber
\end{eqnarray}
Note that the momentum $P^\mu$ is now part of the $8_v$.
By decomposing the representations for each $p$ with respect 
to the common subgroup 
\begin{equation}
SU(8)\supset SO(7)\subset SO(7,1)_{hidden}
\end{equation}
the same sets of $SO(7)$ representations are recovered 
from either (\ref{klist}) or (\ref{olist}). 
This $SO(7)$ already contains one of the hidden dimensions and
classifies the central extensions of types $P,Z,X$ separately as listed in
Table I.

The main point is that the supercharges as well as the central extensions
are now classified under hidden (broken) symmetries of two different types.
The first one $SO(c+1,1)_{hidden}$ relates to 11 or perhaps 12 hidden
dimensions, and the second one $K\subset U$ relates to $U$-duality$.$ The
common compact subgroup $SO(c+1)$ already contains non-perturbative
information about the spacelike hidden dimension, but more information about
the hidden time-like dimension and about $U$-duality is contained in the
larger group structures $K,\,\,SO(c+1,1)$.

\section{Non-perturbative states}

Under the assumption that the superalgebra is valid as a dynamical (broken)
symmetry in the entire theory, all states would belong to multiplets of the
(broken) superalgebra, including the central extensions and the $p$-branes
associated with them. One would then expect to be able to classify the
physical states of the theory according to the (broken) isometries $%
K,\,\,SO(c+1,1)$. However, since these groups are not contained in each
other we should have different modules of $SO(c+1,1)_{hidden}$ and $K\subset
U$ that have intersections with each other in the form of (broken) $SO(c+1)$
multiplets, since this is the largest common subgroup 
\begin{equation}
K\supset SO(c+1)\subset SO(c+1,1)_{hidden}.
\end{equation}
It seems reasonable to make the hypothesis that the complete set of states
of the theory could be classified with either group, but that each such
classification would contain the same set of $SO(c+1)$ representations. One
of our aims is to test this hypothesis. Each one of these classifications
contains non-perturbative states related to either duality or hidden
dimensions. By finding them and studying their couplings consistent with the
superalgebra one would be able to learn certain global properties of the
underlying theory.

Some of the couplings described by $Z_{\mu _1\cdots \mu _p}^{ab}$ are
perturbative while others are non-perturbative in the string language, but
all couplings or states are on an equal footing from the point of view of
the superalgebra and its isometries. One must include open $p$-branes in the
form of $D$-branes since they couple to closed $p$-branes. Therefore, we
expect that various excitations of open/closed charged $p$-branes $X_\mu
^i(\tau ,\sigma _1,\cdots ,\sigma _p)$ (and their supersymmetric partners)
occur on an equal footing in supermultiplets that contain the (broken) group
structures revealed above. String theory states at various excitation levels
by themselves may not necessarily form the needed multiplets in higher
dimensions (10,1) or (10,2) or in $U$-duality. However some combination of
open/closed $p$-brane states are expected to fill complete multiplets of the
isometries or broken symmetries of the global superalgebra. By starting from
the known superstring states, the supermultiplets connected to them can be
found, and the non-perturbative states can be identified.

Following the arguments in \cite{ibyank} the states of the full theory may
be classified as 
\begin{equation}
\phi _{indices}(base)
\end{equation}
where the base consists of the commuting 528 bosonic generators of the
superalgebra. These include the continuous momentum and the quantized
central extensions $Z_{\mu _1\cdots \mu _p}^{ab}$ that are at an equal
footing. These quantum numbers are classified in linear representations of
the (broken) isometries $K$ or $SO(c+1,1)$ as given in Table I 
\footnote{
According to the dimensions of representations in Table I, the 0-brane 
$Z^{ab}$ central extensions seem to correspond to complete linear
representations of $U$ for all dimensions except for $d=3$ ({\bf 120 }is not
a representation $E_{8(8)}).$ Similarly, higher $p$-branes $Z_{\mu _1\cdots
\mu _p}^{ab}$ do not generally form linear representations of $U$.
Furthermore, the $Z_{\mu _1\cdots \mu _p}^{ab}$ seem to form complete
representations of $SO(c,c)$ for all cases except for ($d=5,\,p=3)
$,\thinspace ($d=3,4\,,\,\,p=2)$. We interpret these observations to mean
that the base is not generally a bunch of 
{\it linear} representation of either $T$
or $U$ duality groups, but it is a bunch of 
{\it linear} representation of $K$ or 
$SO(c+1,1).$}. 
If the superalgebra is valid in the full theory then the $%
indices$ must also fall into linear representations of $K$ or $SO(c+1,1)$ in
order to provide a basis for its (broken) isometries. Thus, both the indices
as well as the base contain information about non-perturbative states through
duality transformations or rotations into the hidden dimensions. Classifying
the states under these groups relates the properties of non-perturbative
states to those of the perturbative string states.

A possible scheme for finding the non-perturbative states is as follows.
First identify the perturbative string states, classify them under
supermultiplets and identify their classification under the perturbatively
explicit $SO(c)_L\otimes SO(c)_R.$ Then try to reclassify them under the
bigger (broken symmetry) group $K.$ 
If additional states are needed to make complete $K$
multiplets add them (these extra states are presumably $p$-branes, $D$
-branes). There may be non-unique ways of completing $K$ multiplets. If so,
then try to make it consistent with the presence of the hidden dimensions by
making sure that the $SO(c$+1) representations embedded in $K$ multiplets
are consistent with the structure of the central charges listed in the
table. When this is achieved one should also check that it is all consistent
with a compactification of a collection of states that starts in 11
dimensions, i.e. consistency with 11-dimensional (broken) multiplets with
signature (10,1). One may need to add at this stage more non-perturbative
states that are not in the same $K$-multiplet with some perturbative string
state (presubambly more $p$-or $D$-brane states). So far one should expect
consistency with ``M-theory''. Finally, check if the structure of the
representations that emerge in this way can also be made  consistent with 12
dimensions, with signature (10,2) (perhaps by adding more states). 
In this way many properties of
non-perturbative states could be deduced. Such a program was initiated in
previous papers \cite{ib11d}\cite{ibyank}\cite{sendb}. The results obtained
there (involving string states at many excited levels) are in agreement with
the presence of many of the structures outlined here as far as (10,1) and $K$
structures are concerned. It would be interesting to extend these
ideas to explore (10,2).

It would also be of interest to analyze ``M-theory'' and ``F-theory'' from
the point of view of the general properties of the superalgebra, and
discriminate between general properties based on the superalgebra versus the
properties of the theory that depend on more detailed features. As mentioned
in the footnotes, non-Abelian versions of the superalgebra are possible, and
in fact expected when the $p$-branes propagate on curved backgrounds. It
would be of interest to relate them to properties of various
compactifications of ``M-theory'' and ``F-theory'' in order to learn about
some of their general global properties.


\begin{thebibliography}{99}
\bibitem{dualities}  C. Hull and P. Townsend, Nucl. Phys. B438 (1995) 109.

\bibitem{witten}  E. Witten, Nucl. Phys. B443 (1995) 85, and ``Some comments
on String Dynamics'', hep-th/9507121, to appear in the proceedings of
Strings '95.

\bibitem{ib11d}  I. Bars, Phys. Rev. {\bf D52} (1995) 3567 (=hep-th/9503228 ).

\bibitem{townsend11}  P.K. Townsend, Phys.Lett.{\bf B350} (1995) 184. 
(=hep-th/9501068).

\bibitem{jhs}  J. Schwarz, Phys.Lett.{\bf B367} (1996) 97 (=hep-th/9510086);
hep-th/9509148; ``M-theory extensions of T duality'', hep-th/9601077.

\bibitem{horavawitten}  P. Horava and E. Witten, Nucl. Phys. {\bf B460 (}%
1996) 506 (=hep-th/9510209).

\bibitem{townsend}  P. Townsend, ``p-brane democracy'', hep-th/9507048.

\bibitem{pol}  J. Polchinski, Phys.Rev.Lett. {\bf 75} (1995) 4724 (=
hep-th/9510017). See also, J. Polchinski, S. Chaudhuri, C.V. Johnson ``Notes
on D-branes'', hep-th/9602052.

\bibitem{duffbs}  M. Blencowe and M. Duff, Nucl. Phys. {\bf B310} (1988) 387.

\bibitem{ibjapan}  I. Bars, ``Duality and hidden dimensions'', proceedings
of conference on Frontiers in Quantum Field Theory, Toyonaka, Japan, Dec.
1996.,  USC-96/HEP-B2, to appear (=hep-th/9604nnn).

\bibitem{vafaf}  C. Vafa, ``Evidence for F-theory'', hep-th/9602022; D.R.
Morrison and C. Vafa, ``Compactifications of F-theory on Calabi-Yau
threefolds'' hep-th/9602114, hep-th/9603161; E. Witten, ``Phase transitions
in M theory and F theory'', hep-th/9603150.

\bibitem{martinecf}  D. Kutasov, E. Martinec, and M. O'Loughlin, ``Vaccua of
M theory and N=2 strings'', hep-th/9603116.

\bibitem{ibyank}  I. Bars and S. Yankielowicz, Phys. Rev. {\bf D53} (1996)
4489 ( =hep-th/9511098). I. Bars, ``Consistency between 11D and U-duality'',
hep-th/9601164 .

\bibitem{sendb}  A. Sen, Nucl. Phys. B450 (1995) 103.

\bibitem{haag}  R. Haag, J. Lopuszanski and M. Sohnius, Nucl.Phys.{\bf B88 }$%
\left( 1975\right) $ 257.

\bibitem{sezgin}  For some examples see: E. Bergshoeff and E. Sezgin, Phys.
Lett. {\bf B232 }(1989) 96; {\bf B354} (1995) 256 (=hep-th/9504140).
\end{thebibliography}
\end{document}